\documentstyle[12pt,epsf]{article}
\def\be{\begin{equation}}
\def\ee{\end{equation}}
\def\bea{\begin{eqnarray}}
\def\eea{\end{eqnarray}}
\def\g{\gamma}
\def\p{\partial}
\def\t{\tilde}
\def\G{\Gamma}
\begin{document}

\begin{titlepage}
\bigskip
\rightline{}
\rightline{hep-th/0207270}
\bigskip\bigskip\bigskip\bigskip
\centerline{\Large \bf {Colliding Kaluza-Klein Bubbles}}
     \bigskip\bigskip
      \bigskip\bigskip

  \centerline{\large Gary T. Horowitz and Kengo Maeda}
     \bigskip\bigskip
  \centerline{\em Department of Physics, UCSB, Santa Barbara, CA 93106}
  \centerline{gary@physics.ucsb.edu, maeda@cosmic.physics.ucsb.edu}
	    \bigskip\bigskip

\begin{abstract}
 We construct an exact solution describing the collision of two Kaluza-Klein
``bubbles of nothing" in 3+1 dimensions. When the bubbles collide, a
curvature singularity forms which is hidden inside an event horizon.
However, unlike  the formation of ordinary black holes,  in this case
the spacetime resembles the entire maximally extended Schwarzschild solution.
We also point out that there are inequivalent bubbles
that can be constructed from Kerr black holes.
\end{abstract}
\end{titlepage}

 \baselineskip=18pt

\setcounter{equation}{0}
 \section{Introduction}
Witten showed twenty years ago that the standard (nonsupersymmetric)
Kaluza-Klein vacuum,
Minkowski spacetime cross a circle, was unstable to nucleating a ``bubble
of nothing"~\cite{Witten}. 
The resulting spacetime only exists outside a minimal area
sphere which accelerates out to null infinity. At large radius,
the solution approaches the standard Kaluza-Klein vacuum 
and has zero total mass.
Various generalizations of this bubble spacetime have since been
found, and have been extensively studied.
They arise
as possible end-states of the quantum decay of various
configurations including magnetic flux tubes, and flux 
branes~\cite{Dowker:1995gb, Gutperle:2001mb}.
They have also been studied as
interesting spacetimes in their own right~\cite{clean, Birmingham:2002st},
as they are
some of the few explicitly known
examples of nonsingular, time dependent solutions of 
Einstein's equation. 

If a spacetime is unstable to nucleating one bubble, it may be able to
nucleate a second one far from the first. The natural question is then what
happens when the two bubbles collide. Since the spacetime only exists on
one side of the bubble, this is quite different from the usual collision
of bubbles of true vacuum inside a false vacuum
in field theory\footnote{It is perhaps more
analogous to the collision of ``end of the world" branes \cite{Fabinger:2000jd}.}.
In that case, one can approximate
the collision of two widely separated bubbles by  the collision of flat domain
walls and show that black holes do not form \cite{Moss:iq}.

We will show that the collision of
two Kaluza-Klein bubbles results in a spacelike curvature
singularity. This singularity is hidden behind an event horizon, so one
indeed forms a black hole. However, there are two qualitative differences
from the usual process of black hole formation. First, (part of) the spacetime
resembles the maximally extended Schwarzschild solution, including
the white hole singularity and the second asymptotically flat region.
Second, the singularity and event horizon 
extend all the way out to null infinity.

To obtain these results, we construct and study an exact solution describing
the collision of  two bubbles. Our solution is not in $4+1$ dimensions, the 
usual context for Kaluza-Klein theory, but rather in $3+1$ dimensions.
This is not a serious limitation since $3+1$ dimensional general relativity,
with one direction compactified at infinity, has vacuum solutions which are
identical to the higher dimensional bubbles. This is because
the usual  Kaluza-Klein bubble is obtained by analytic continuation of
the five dimensional euclidean Schwarzschild solution. By taking the
same analytic continuation of the four dimensional euclidean black hole, one
obtains a 3+1 bubble solution with the same properties
as the higher dimensional one.

It is difficult to study the collision of Kaluza-Klein bubbles in 4+1
dimensions, but in 3+1 dimensions, it is straightforward to construct
an exact solution.
This is because the spacetime describing two static black holes in 3+1
dimensions is axisymmetric, and all static axisymmetric vacuum solutions
can be found by solving a linear equation. These are known as the Weyl metrics.
(See \cite{Emparan:2001wk} for a recent discussion.)
The spacetime describing two static Schwarzschild black holes has a conical
singularity on the axis between them. This strut provides the force to
balance the gravitational attraction of the black holes.
However, after the 
analytic continuation, the strut is removed.
On an initial time slice,
the spacetime describes two bubbles and is completely nonsingular. Since
the exact double bubble
solution is known, one can now follow the evolution and see
the formation of curvature singularities 
when the bubbles collide. We expect that the collision 
of higher dimensional bubbles will be similar.

One can also construct bubble solutions by analytic continuation of
Kerr black holes. We briefly discuss the collision of these Kerr bubbles.
An exact solution can again be constructed since the spacetime describing
two Kerr black holes with their spins parallel and aligned along the symmetry 
axis is known. Unfortunately, the natural analytic continuation from the
black holes to the bubbles produces a solution with closed timelike curves at 
infinity. This is puzzling since a previous analytic continuation of
a single Kerr black hole to a Kerr bubble did not have this problem 
\cite{clean}. We resolve
this puzzle by showing that there are inequivalent Kerr bubbles that can
be constructed from the same Kerr black hole. This is not the case for the
bubbles constructed from a Schwarzschild black hole, and is a direct result
of the reduced symmetry of the spacetime.

\setcounter{equation}{0}
\section{Preliminaries}
\subsection{Review of single bubble}
We begin by reviewing the properties  of a single Kaluza-Klein bubble in $3+1$
dimensions.
The euclidean Schwarzschild metric is
\be\label{euclbh}
ds^2 = \left[1-{2M\over r}\right] d\chi^2 +
\left[1-{2M\over r}\right]^{-1} dr^2 + r^2(d\theta^2 + \sin^2\theta d\phi^2)
\ee
where $r\ge 2M$ and regularity at $r=2M$ requires $\chi$ to be periodic 
with period $8\pi M$.
We wish to analytically continue the two-sphere metric into two dimensional
de Sitter space. 
This can be done globally by setting $\theta-\pi/2 = i\tau$,
so the metric becomes
\be
ds^2 = \left[1-{2M\over r}\right] d\chi^2 +
\left[1-{2M\over r}\right]^{-1} dr^2 + r^2(-d\tau^2 + \cosh^2\tau  d\phi^2).
\ee
The geometry at $\tau=0$ looks like $R^2$ with a disk of radius $2M$ removed.
There is a circle over every point (parameterized by $\chi$)
which has constant radius at infinity,
but smoothly goes to zero at $r=2M$. So $r=2M$ is not a boundary, but it
is a circle of minimal length. This is the ``bubble of nothing". As one
moves away from $\tau=0$, the bubble grows exponentially. The induced metric
on the bubble is just two dimensional de Sitter spacetime.

It was shown in~\cite{clean} that 
this spacetime has observer dependent horizons 
analogous to de Sitter spacetime. The Penrose diagram is shown in Fig 1.
The generators of null infinity
are incomplete, and observers at different points on the $\phi$
circle lose causal
contact at late time. 
%%%%%%%%%%%%%%%%%%%%%%%%%%%%%%%
\begin{figure}[htb]
     \centerline{\epsfxsize=9.6cm 
       {\epsfysize=12.5cm
             \epsffile{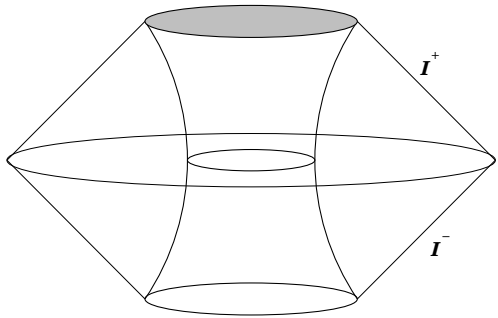}}}
      \caption{The Penrose diagram of a single bubble. There is no spacetime
      inside. Future and past null infinity are incomplete since they are
      cut-off by the bubble.}
\label{7eps}
\end{figure}
%%%%%%%%%%%%%%%%%%%%%%%%%%%%%%%%
Note that the minimum size of the bubble (at $\tau=0$) is related to the size 
of the Kaluza-Klein circle at infinity. Thus, in order to place two bubbles 
in the same asymptotic spacetime, they must have the same minimum size.
This translates into starting with two static black holes of  equal mass.
In the two black hole
solution that we review in the next section,
 we will not have the analog of $\theta$
to analytically continue and obtain global coordinates. Instead, since
axisymmetry will be manifest, we will do the analog of setting
$\phi = it$ in (\ref{euclbh}).
This is like writing de Sitter in static coordinates. The timelike
Killing field becomes null on the surfaces $\theta=0,\pi$.
These are the de Sitter horizons. Actually, in the bubble solution, $\p/\p t$ 
is analogous to a boost symmetry, and the horizons are more analogous to
Rindler horizons. On the other side of these horizons,
$\p/\p t$ is spacelike.

\subsection{Two black hole solution}

We now review the solution for two static black holes~\cite{Myers}. 
Every static axisymmetric vacuum solution can be written in the form
\be\label{weyl}
ds^2 = -e^{2\Phi} d\hat t^2 + e^{2(\g -\Phi)} (dr^2 + dz^2) +
 e^{-2\Phi}r^2d\phi^2
\ee
where $\Phi$ is an axisymmetric solution of the flat space Laplace equation and
$\g$ is determined in terms of $\Phi$. If $\Phi$
is the Newtonian potential for a
rod of length $2M$ and mass $M$, the resulting metric is the Schwarzschild
black hole. The solution with two equal mass
black holes is obtained by letting $\Phi$ be
the Newtonian potential for two such rods separated by $2L$ with $L>M$.
It will be convenient to work with the combinations $a=L+M,\ b=L-M$, so
$a>b>0$.
Setting
$$\rho_1 = [r^2 + (z-a)^2 ]^{1/2}, \qquad
\rho_2 = [r^2 + (z+b)^2 ]^{1/2}  $$
\be\label{defrho}
\t \rho_1 = [r^2 + (z-b)^2 ]^{1/2}, \qquad
\t \rho_2 = [r^2 + (z+a)^2 ]^{1/2}
\ee
the solution for $\Phi$ is
\be\label{Phi}
e^{2\Phi}=\frac{\rho_1+\tilde{\rho_1}+b-a}
{\rho_1+\tilde{\rho_1}+a-b}\times
\frac{\rho_2+\tilde{\rho_2}+b-a}
{\rho_2+\tilde{\rho_2}+a-b}.
\ee
The corresponding solution for $\g$ is
\be\label{gamma}
e^{2\g} =\G_1 \G_2 \G_3 \G_4
\ee
where
\be\label{g1}
\G_1 = {\rho_1 \t \rho_1 + (z-a)(z-b) +r^2\over 2\rho_1 \t \rho_1}, 
\ee
\be\label{g2}
\G_2 = {\rho_2 \t \rho_2 + (z+a)(z+b) +r^2\over 2\rho_2 \t \rho_2}, 
\ee
\be\label{g3}
\G_3 = {\rho_1 \t \rho_2+ z^2-a^2 + r^2\over
         \rho_1 \rho_2 +(z-a)(z+b) +r^2}, 
\ee
\be\label{g4}
\G_4 = 	{\t \rho_1 \rho_2 +z^2-b^2 + r^2 \over
	 \t \rho_1 \t \rho_2 +(z+a)(z-b) + r^2}. 
\ee
The black hole horizons occur where $e^{2\Phi}$ vanishes. This implies either
$\rho_1 + \t \rho_1 = a-b$ or $\rho_2 + \t \rho_2 = a-b$. The first condition
is satisfied when $r=0$ and $b\le z \le a$ and the second  when
$r=0$ and $-a\le z \le -b$. (These are just the locations of the Newtonian
rods.)

 To check regularity on the rotation axis $r=0$, $|z|>a$ or $|z|<b$,
let us calculate the following quantity 
\be
\alpha=\lim_{r\to 0}\frac{\sqrt{g_{\phi\phi}}\, \Delta \phi}
{\int^r_0 \sqrt{g_{rr}}dr},
\ee
where $\Delta \phi$ is the period of $\phi$.
Regularity on the axis requires
$\alpha=2\pi$. From the metric~(\ref{weyl}),
one finds
\be
\alpha=e^{-\gamma}\Delta \phi,
\ee
where $\gamma$ is evaluated at $r=0$. From (\ref{g1})-(\ref{g4}), one finds
\be
e^\gamma=1
\ee
for $|z|>a$ and
\be
e^\gamma=\frac{4ab}{(a+b)^2}
\ee
for $|z|<b$.
It follows that one can choose $\Delta \phi$ to remove the conical singularity
either between the black holes or from each black hole to infinity,
but not both. (This was first shown in \cite{Einstein}.)
Physically, these singularities
are struts required to hold the black holes apart. They will be absent in
the two bubble solution we  construct next.

\setcounter{equation}{0}
\section{Colliding bubble solution}

\subsection{Analytic continuation}

The colliding bubble solution is obtained by a double analytic continuation
of the above two black hole solution. We first
set $\hat t=i\chi$ in (\ref{weyl}) to obtain the euclidean solution.
Expanding the metric near the
horizons, one sees that this can be done without introducing additional
conical singularities if $\chi$ is periodic with period
\be
 \Delta \chi = {8\pi a(a-b)\over (a+b)}.
\ee
(If one had started with two black holes of different mass, this would not
have been possible.)
We then set $\phi = it$ to obtain  the double bubble solution:
\be\label{doublebub}
ds^2=e^{2\Phi} d\chi^2+e^{2(\g-\Phi)}(dr^2 + dz^2) -e^{-2\Phi}r^2dt^2.
\ee
  This spacetime looks static,
but as we mentioned above, these coordinates do not cover the entire spacetime.
$\p/\p t$ is like a boost symmetry and there are Rindler horizons at $r=0$.
The collision of the two bubbles takes place in the time dependent region
beyond the Rindler horizon between the two bubbles.

We will explore this region of the spacetime in the next subsection,
but first consider the time symmetric
surface $t=0$. This is completely smooth with no conical singularities.
The previous struts arose because of the periodic identification of $\phi$,
but we have analytically continued $\phi$ to obtain our new time coordinate.
The two previous spherical event horizons become minimal circles which are 
the two initial 
bubbles. It might appear that the horizons become line segments 
$r=0$, $b\le |z| \le a$. However, this is like the static patch of de Sitter.
One
needs two copies of the metric to cover the entire constant $t$ surface.
The axis $r=0$ between the two black holes  becomes a two-sphere $\Sigma$
linking the two bubbles. This is because for each $-b \le z \le b$, there is
now a circle parameterized by $\chi$ which smoothly shrinks to zero size at 
$z=\pm b$. 
Since the metric involves only even powers of $r$, one can smoothly extend
the metric past $\Sigma$ by letting $r$ take negative values. Letting $r$ and
$z$ take all real values, the metric (\ref{doublebub}) covers the entire
constant $t$ 
surface\footnote{There is a subtlety at the bubbles $b\le |z| \le a$.
The limit $r\rightarrow 0$ with $r>0$, is different from the limit
$r\rightarrow 0$ with $r<0$. These two line segments combine to form the
circle which is the bubble.}. Since the space has a reflection symmetry,
$r\rightarrow -r$,
$\Sigma$ is a minimal two-sphere.
This two-sphere will play an important role in what follows.

\subsection{Bubble collision and black hole formation}

The static coordinates we have introduced in (\ref{doublebub}), even
with $r$ extended to negative values, only cover the region of
spacetime which is spacelike related to the two-sphere $\Sigma$. To describe
the spacetime to the future of $\Sigma$ we must
analytically continue the metric past the 
Rindler horizon at $r=0$. The simplest way to do this, is 
to set $r=i\tilde{r}$. The metric remains real   since it
does not include any odd functions of $r$. Just like the
region inside the horizon of Schwarzschild, the coordinate $t$ now becomes
spacelike, and $\t r$ is the new time coordinate.
Since there are two commuting spacelike killing vectors,
$\partial/\p t$ and $\partial/\p \chi$, it is convenient to
introduce double null coordinates

%%%%%%%%%%%%%%%%%%%%%%%%%%%%%%%%%%%%%%%%%%%%%%%%%%%%%%
\be
%%%%%%%%%%%%%%%%%%%%%%%%%%%%%
\label{null-coordinate}
%%%%%%%%%%%%%%%%%%%%%%%%%%%%%
u=\tilde{r}+z, \qquad v=\tilde{r}-z,
\ee
so the resulting metric is
\be\label{double1}
ds^2=e^{2\Phi} d\chi^2+e^{-2\Phi}\tilde{r}^2dt^2-e^{2(\g-\Phi)}dudv.
\ee
The functions $\rho_i, \t \rho_i$ now take the form
$$ \rho_1=[(a-u)(v+a)]^{1/2}, \qquad \rho_2=[(u+b)(b-v)]^{1/2} $$
\be\label{trho_1}
\tilde{\rho_1}=[(b-u)(v+b)]^{1/2}, \qquad \tilde{\rho_2}=[(u+a)(a-v)]^{1/2}. 
\ee
As easily checked, the
metric~(\ref{double1}) first becomes singular on the following two null
hypersurfaces
\be\label{real}
u=b, \qquad \mbox{or}\qquad  v=b
\ee
since both $\G_1$ and $\G_2$ diverge.
However this is just a coordinate singularity which can be removed by
introducing new double null coordinates:
\be\label{new-null-coordinate}
U=-\sqrt{b-u}, \qquad V=-\sqrt{b-v}.
\ee
Inverting these relations, $u$ and $v$ become functions of $U$ and $V$,
respectively 
\be\label{uv}
u(U)=b-U^2, \qquad v(V)=b-V^2.
\ee

The corresponding new double null metric is
\be\label{double2}
ds^2=e^{2\Phi} d\chi^2+e^{-2\Phi}\tilde{r}^2dt^2-
e^{2(\tilde{\g}-\Phi)}dUdV,
\ee
where
\be
e^{2\tilde\g}=\tilde\G_1\tilde\G_2\tilde\G_3\tilde\G_4
\ee
and
\be
\tilde\G_1=\frac{\rho_1\tilde{\rho}_1-uv
-\frac{a+b}{2}(u-v)+ab}
{\rho_1\sqrt{2b-V^2}},
\ee
\be
\tilde\G_2=\frac{\rho_2\tilde{\rho}_2-uv
+\frac{a+b}{2}(u-v)+ab}
{\t \rho_2\sqrt{2b-U^2}},
\ee
\be
\tilde\G_3=\frac{\rho_1\tilde{\rho}_2-uv-a^2}
{\rho_1\rho_2-uv-\frac{a-b}{2}(u-v)-ab},
\ee
\be
\tilde\G_4=\frac{\tilde{\rho}_1\rho_2-uv-b^2}
{\tilde{\rho_1}\tilde{\rho_2}-uv+\frac{a-b}{2}(u-v)-ab}.
\ee
 
We now check for singularities. It turns out that some
metric components diverge (or
vanish)
along the circle
\be\label{circle}
U^2+V^2=2b \qquad (u+v=0).
\ee
This circle can be divided into two timelike segments ($UV<0$) and two
spacelike segments ($UV>0$).
We now show that the metric remains smooth on the timelike segments, 
which correspond to the two bubbles approaching each other. However, the
curvature diverges on the future spacelike segment  which corresponds
to the collision of the two bubbles. The past spacelike segment
is just the two-sphere $\Sigma$ between the two bubbles at the moment of
time symmetry $\t r=0$.

Along a null hypersurface $V=V_0>0$, one can find the asymptotic behavior
of the metric near the future spacelike segment as
\be\label{asy}g_{\chi\chi}\sim O(\epsilon^4), \qquad
g_{tt}\sim O(\epsilon^{-2}),  \qquad
-g_{UV}\sim O(\epsilon^4)
\ee
where $\epsilon$ is a small positive number defined as
\be\label{u-epsilon}
U=\sqrt{2b-V_0^2}-\epsilon.
\ee
% figure ----------------------
(The same asymptotic behavior will be found along a null
hypersurface $U=U_0>0$ because of symmetry under $U\leftrightarrow V$.) 
Since the norm of the Killing field $\p/\p t$ is diverging, this is a strong
indication of a curvature singularity. One need only check that $\epsilon=0$
can be reached in finite affine parameter. But this is true, 
since the affine parameter $\lambda$ of the
null geodesic along $V=V_0$ is proportional to
$\int g_{UV}dU$. Alternatively, consider a congruence of null geodesics
with each geodesic following a curve of constant $t, \chi, V$.
Since $g_{tt}$ diverges in a finite
affine length, while $g_{\chi\chi}$ shrinks to zero, the null
congruence experiences infinite distortion indicating a strong curvature
singularity.

One can see that this curvature singularity looks like the Schwarzschild
singularity as follows.
The Schwarzschild metric near the singularity becomes
\be\label{Sch1}
ds^2\simeq
\frac{2m}{r}dt^2-\frac{r}{2m}dr^2+r^2(d\theta^2+\sin^2\theta d\phi^2),
\ee
where $m$ is a positive mass parameter.
After a coordinate transformation $r=\epsilon^2$, the metric
becomes
\be\label{Sch2}
ds^2\simeq
\frac{2m}{\epsilon^2}dt^2
-\frac{2\epsilon^4}{m}d\epsilon^2+
\epsilon^4(d\theta^2+\sin^2\theta d\phi^2).
\ee
Introducing double null coordinates $U=\sqrt{2/m}\ \epsilon+\theta$
and $V=\sqrt{2/m}\ \epsilon-\theta$, and replacing $\phi$ with $\chi$,
one obtains same asymptotic behavior as Eq.~(\ref{asy}).

For the timelike segment~($V<0$) of the circle (\ref{circle}),
each metric component behaves as\footnote{A
key difference between this and the future 
spacelike segment is that both factors
in $e^{2\Phi}$ vanish on the spacelike segment, while only one vanishes
on the timelike segment.}
\be
g_{\chi\chi}\sim O(\epsilon^2), \qquad
g_{tt}\sim O(1), \qquad
-g_{UV}\sim O(1)
\ee
%%%%%%%%%%%%%%%%%%%%%%%%%%%%%%%
\begin{figure}[htb]
 \centerline{
            {\epsfxsize=9.0cm 
             \epsfysize=11.5cm        
            \epsffile{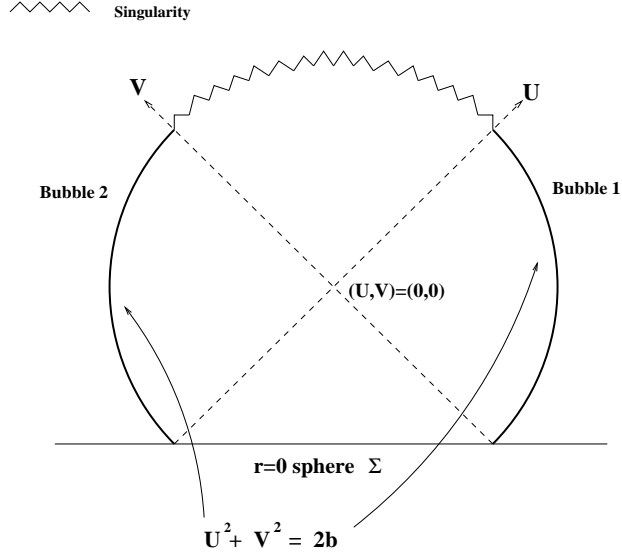}}}
      \caption{The spacetime between the bubbles in $U,V$ coordinates.
      The fact that the bubbles appear to move apart initially is a coordinate
      artifact -- they actually start at rest. 
 The resulting singularity looks like 
the Schwarzschild singularity. }
\label{1eps}
\end{figure}
%%%%%%%%%%%%%%%%%%%%%%%%%%%%%%%
along the null geodesic $V=V_0<0$.
This looks like a flat geometry near the axis of the following
cylindrical coordinates
\be\label{cyl}
ds^2\sim dt^2+\epsilon^2d\chi^2+d\epsilon^2-d\tilde{r}^2.
\ee
This implies that the timelike segment is completely regular
and represents an orbit of one of the colliding bubbles. Therefore,
the two bubbles collide with each other at the null
points ($U=\sqrt{2b},\,V=0$) and 
($V=\sqrt{2b},\,U=0$)~(See Fig.~\ref{1eps}). This is metrically
at zero proper separation since $g_{UV} \rightarrow 0$. At this point, a
Schwarzschild type strong curvature singularity appears.

Since every future-directed causal curve in the region
$\tilde{r}>0$ and $|z|\le b$ terminates at the singularity,
the Rindler horizon $\tilde{r}=0$ between the two  bubbles
corresponds to a {\it black hole event horizon}. This is a Killing horizon
since the Killing field $\p/\p t$ is timelike outside,  null on the 
horizon, and spacelike inside. 
The black hole is compact and has topology $S^2$.
The two-sphere $\Sigma$ linking the two bubbles at $t=0$ is the
bifurcation two-sphere. This structure is very similar to the maximally
extended Schwarzschild
solution except, of course, that the horizon is not spherically symmetric.
It is inhomogeneous along the $z$ direction. It is noteworthy that 
this inhomogeneity 
does not grow infinitely toward the singularity since
the ratio $g_{\chi\chi}/g_{UV}\sim O(1)$.
The horizon area $A$ is easily calculated to be:
\be\label{area}
A=\frac{2^6\pi (a-b)\,a^2\,b^2}{(a+b)^3}.
\ee
%%%%%%%%%%%%%%%%%%%%%%%%%%%%%%%
\begin{figure}[htb]
     \centerline{{\epsfxsize=9.8cm 
       \epsfysize=12.6cm    
       \epsffile{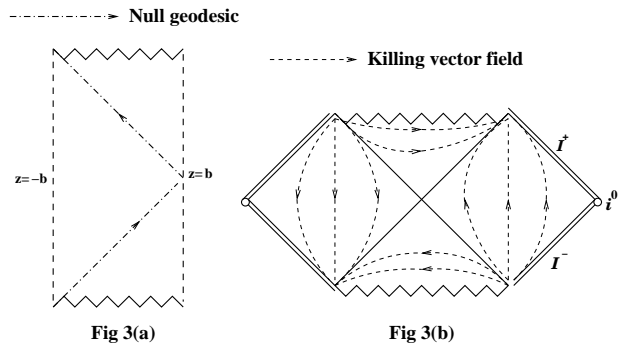}}}
      \caption{(3a) The evolution of the two-sphere $\Sigma$ looks like
      a closed dust-filled Friedman universe. (3b) The geometry on
      the surface $z=0$ half-way between the two bubbles looks like
      a Schwarzschild black hole, except that null infinity is incomplete.}
\label{5eps}
\end{figure}
%%%%%%%%%%%%%%%%%%%%%%%%%%%%%%%

To visualize the geometry between the two bubbles, it is convenient to 
first fix $t$, and consider the evolution of the two-sphere $\Sigma$ linking
the two bubbles (recall that $t$ is a spacelike coordinate in this region).
This resembles a closed universe that collapses to a curvature singularity
in the future and past\footnote{By time reflection symmetry, 
the evolution of the
two bubbles to the past
must be identical to their evolution in the future discussed above.}.
In fact, the Penrose diagram is identical to a
dust filled Friedmann solution (see Fig.~\ref{5eps}) 
since a light ray starting from a bubble when the
bubbles collide in the past, can make it exactly once around before the
bubbles collide in the future. Alternatively, one can fix $\chi$ and set
$z=0$. This corresponds to the plane exactly halfway between the bubbles.
The Penrose diagram of the two dimensional metric induced on
this surface looks exactly like the maximally extended Schwarzschild
solution, including the white hole singularity and the second
asymptotically flat region! Although the causal 
structure is identical, there is an important difference. In Schwarzschild,
although it looks like the singularity is touching null infinity, this is
an artifact of the conformal rescaling. Metrically, the singularity
is contained inside a sphere of constant radius. However, in the bubble 
collision, the singularity really extends all the way to null infinity.

One can estimate how long it takes for the two bubbles to collide as
follows. This time is the proper time along the geodesic $z=0$,
$t,\chi$ constant, from $\t r=0$ to the singularity. 
Suppose the separation between the two bubbles is much greater 
than their size, $b\gg a-b$. Then  by rescaling the variables in
(\ref{double2}) by appropriate powers of $b$ depending on their dimension,
one can  show that the time to the singularity is proportional to $b$.
Similarly, the proper distance between the bubbles is also proportional 
to $b$. So the time to the collision is proportional to their separation.
This is similar to the answer one would expect by simply considering two
hyperboloids in Minkowski spacetime. The curved spacetime around the bubbles
does not significantly affect their time to collide.

\subsection{Asymptotic structure}

Now let us turn to the spacetime between the asymptotically flat region 
and the first bubble: $|z|\ge a$. 
Because of the symmetry $z\leftrightarrow -z$,
we can consider only the $z\ge a$ region without loss of generality.
There is again a Rindler horizon at
$r=0$ which we can extend beyond by setting $r = i\tilde r$, and
introducing the double null coordinates $u,v$ as in (\ref{null-coordinate}). 
As in the region $|z| \le b$, there is a coordinate singularity
at the null hypersurface $v=-a$, where $\G_1$ diverges.
To avoid this singularity one can introduce a new
null coordinate $V$ as
\be\label{new-V}
V=-\sqrt{-v-a}.
\ee
The corresponding double null metric is
\be\label{double22}
ds^2=e^{2\Phi} d\chi^2+e^{-2\Phi}\tilde{r}^2dt^2-
e^{2(\bar{\g}-\Phi)}dudV,
\ee
where
\be
e^{2\bar\g}=\bar\G_1\bar\G_2\bar\G_3\bar\G_4,
\ee
and
\be
\bar\G_1=\frac{\rho_1\tilde{\rho}_1-uv
-\frac{a+b}{2}(u-v)+ab}
{\tilde{\rho}_1\sqrt{u-a}},
\ee
\be
\bar\G_2=\frac{\rho_2\tilde{\rho}_2-uv
+\frac{a+b}{2}(u-v)+ab}
{2\rho_2\tilde{\rho}_2},
\ee
\be
\bar\G_3=\tilde{\G}_3
\ee
\be
\bar\G_4=\tilde{\G}_4
\ee
The metric components are now singular along
the following parabolic timelike orbit:
\be\label{sphere}
u=V^2+a \quad (u+v=0).
\ee
One can see that this orbit corresponds to the nonsingular motion of
the bubble on right hand side as follows.
Along the null geodesic $u=u_0>a$, the asymptotic
behavior of the metric~(\ref{double22}) is
\be
g_{\chi\chi}\sim O(\epsilon^2),
\ee
\be
g_{tt}\sim O(1),
\ee
\be
-g_{uV}\sim O(1),
\ee
where $\epsilon$ is defined as
\be
V=\sqrt{u-a}-\epsilon.
\ee
This is same as the behavior of the metric~(\ref{cyl}).
Therefore, the spacetime has no singularity beyond the Rindler horizons
outside the two bubbles~($|z|\ge a$).

Finally, let us investigate null infinity of this spacetime.
Since the double bubble 
metric~(\ref{doublebub}) does not cover the whole spacetime,
null infinity is divided into two regions: (i) null infinity
in $r>0$, (ii) null infinity in $\tilde{r}>0$ and $|z|>a$.
In case (i), the geometry of the solution approaches
\be
ds^2\simeq d\chi^2+dr^2+dz^2-r^2dt^2
\ee
as $r\to \infty$.
Therefore, these coordinates only cover null infinity up to the
Rindler horizons at $r=0$.
In case (ii), the $uV$ component of the metric~(\ref{double22})
becomes
$$
g_{uV}=-\frac{1}{4}\cdot
\frac{V^2+\frac{a-b}{2}-V\sqrt{V^2+a-b}}
{\sqrt{V^2+a-b}}\cdot
\frac{\sqrt{(V^2+a+b)(V^2+2a)}+V^2+\frac{3a+b}{2}}
{\sqrt{(V^2+a+b)(V^2+2a)}}
$$
\be\label{uV}
\times
\frac{a+V^2-V\sqrt{V^2+2a}}{V^2-V\sqrt{V^2+a+b}+\frac{a+b}{2}}
\cdot
\frac{\sqrt{(V^2+a-b)(V^2+a+b)}+a+V^2}
{\sqrt{(V^2+2a)(V^2+a-b)}+V^2+\frac{3a-b}{2}}
\ee
in the $u\to \infty$ limit. So, one can easily find
\be
g_{uV}\simeq -\frac{a^2(a-b)^2}{4(a+b)^2V^3}
\ee
as $V\to \infty$. Since the affine parameter along the null
geodesic of the null infinity is given by $-\int g_{uV}dV$,
the null infinity is future incomplete.

It is not surprising that null infinity is incomplete, since that is also
true for a single bubble spacetime. In fact, one can describe the
asymptotic structure as follows.
Each bubble intersects
null infinity in a circle, and the two circles intersect in two points $p$ and
$q$ (see Fig. 4).
The curvature singularity approaches these two points. The black hole
event horizon is the boundary of the past of these two points on future
null infinity.
%%%%%%%%%%%%%%%%%%%%%%%%%%%%%%%
\begin{figure}[htb]
 \centerline{
     {\epsfxsize=10.0cm 
      \epsfysize=12.0cm     
      \epsfbox{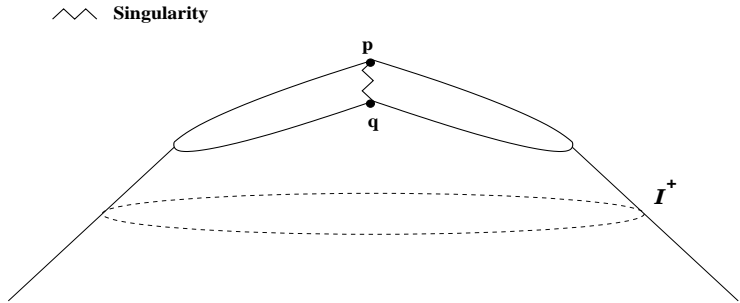}}}
      \caption{The structure of null infinity in the colliding bubble 
      solution. The singularity is in the interior of the spacetime, but
      approaches null infinity asymptotically.}
\label{8eps}
\end{figure}
%%%%%%%%%%%%%%%%%%%%%%%%%%%%%%%%
\setcounter{equation}{0}
\section{Colliding Kerr bubbles?}

One can start with the (four dimensional)
Kerr metric and do a double analytic continuation to
obtain another bubble solution. This bubble starts to expand exponentially
(like the bubble obtained from the Schwarzschild black hole) but then
it stops expanding \cite{clean}.
Null infinity is complete. One might ask what happens
when two of these Kerr bubbles collide. One might expect that if the
bubbles collide when they are both expanding exponentially, then the solution
will resemble the collision of Schwarzschild bubbles with a spacelike
curvature singularity. But what happens when the bubbles stop expanding?
Does the singularity become timelike? Is cosmic censorship violated?

To try to address this,
one can start with a stationary, axisymmetric solution
consisting of two Kerr black holes with their spins either parallel or
anti-parallel. Since there are still two commuting Killing fields,
this solution is again known explicitly \cite{kerr}.
As before, there is a strut keeping the black holes apart.
When one does a double analytic continuation, the strut again disappears
and one obtains a spacetime
with two colliding Kerr bubbles. Surprisingly, one finds that the bubbles
continue to expand exponentially. They do not stop.

The problem is that the analytic continuation needed here
$\phi\rightarrow it$ is different from the one used in~\cite{clean}  
 $\theta \rightarrow
\pi/2 + i\tau$. For a Schwarzschild bubble, the resulting metrics are
equivalent. For a Kerr bubble, the resulting metrics are different. In
other words, there are really two solutions describing
Kerr bubbles. This is possible even
though they both are obtained by analytic continuation from the same
euclidean solution. As a simple example, consider the distorted two-sphere
\be\label{example}
ds^2 = d\theta^2 + (\sin^2 \theta + c^2 \sin^4\theta)d\phi^2. 
\ee
If one sets $\phi=it$, one obtains the Lorentzian space
\be
ds^2_1 = -(\sin^2 \theta + c^2 \sin^4\theta)dt^2 +d\theta^2. 
\ee
These coordinates do not cover the entire spacetime.
There is a boost symmetry analogous to de Sitter space and $\theta =0,\pi$
are horizons. Indeed, when
$c=0$ this is precisely de Sitter space in static coordinates. So we know
that to complete the spacetime we should really take two copies of this
coordinate patch and glue them together at $\theta=0,\pi$. The resulting
spacetime has a circle of minimal length $2\pi$.
If one instead sets $\theta =\pi/2 + i\tau$ in (\ref{example}),
one obtains the Lorentzian space
\be
ds^2_2 = -d\tau^2 + (\cosh^2 \tau + c^2 \cosh^4 \tau)d\phi^2. 
\ee
This describes an axisymmetric spacetime with a circle that contracts down
to a minimum size $2\pi(1+c^2)^{1/2}$ and then expands. Since the
circles have different minimal lengths, the spacetimes are clearly
inequivalent. (The symmetries of
the two spacetimes are also different.)

Similarly, the two Kerr bubbles are inequivalent.
However, only one is physically realistic. If one sets $\phi=it$ in
euclidean Kerr,
one finds closed timelike curves asymptotically. This is because the
identification that is needed to obtain a smooth euclidean solution
involves a rotation of $\phi$ together with a shift in imaginary time.
The rotation dominates at large radius, so if $\phi$ becomes a new time
coordinate, these closed circles become timelike. So the
solution one obtains by starting with two stationary Kerr black holes
and analytically continuing $t=i\chi, \phi = it$ has closed
timelike curves asymptotically, and is not the one we want.
This problem does not
arise if one analytically continues $\theta$ in the single Kerr metric
since then
$\phi$ remains spacelike.
In that case, the Kaluza-Klein circles just open up asymptotically.
Unfortunately, there is no analogous analytic continuation in the two
Kerr black hole solution. In fact, since the circles open up,
there may not exist a solution with two Kerr bubbles.
Indeed, it has been argued that in a spacetime with one Kerr bubble, one
cannot nucleate another~\cite{clean}.

In more than four dimensions, rotating black holes have more than one angular
momentum parameter. If some of these are zero, one has spherical symmetry
in the  directions orthogonal to the planes of rotation. One can construct
a bubble solution by analytically continuing time and one of the angles
of spherical symmetry. This will not produce closed timelike curves, 
since the identification required for a smooth euclidean solution only
involves angles in the planes of rotation. 

\setcounter{equation}{0}
\section{Discussion}

We have  shown that when two Kaluza-Klein bubbles collide, they produce a 
spacelike curvature singularity which is similar to the one inside a
Schwarzschild black hole. There is also an event horizon enclosing the
singularity. This absolute event horizon is quite different from the 
observer dependent horizons that are present even
in the single bubble spacetime. It
shares many properties of a standard black hole horizon, e.g. it is the
boundary of the past of null infinity. However, it also shares properties
of a Rindler horizon associated with an accelerated observer, since
the event horizon extends out and touches null infinity. 

Our entire analysis has been in four spacetime dimensions.
Although we do not have an exact solution in higher dimensions, we expect
the collision of two bubbles to again produce a curvature singularity. 
There is, however, an important difference from the four dimensional case.
To see this, let us consider the symmetries. The five dimensional
Schwarzschild black hole has $SO(4) \times R$ isometries. This becomes
$SO(4)\times U(1)$ for the euclidean black hole, and $SO(3,1) \times U(1)$
for the single bubble. Adding a second bubble breaks this symmetry down to 
$SO(2,1) \times U(1)$. (Note that a similar argument in four dimensions
shows that the two bubble solution has $SO(1,1) \times U(1)$ isometries
which are just the two commuting Killing fields we have used above.)
In five dimensions, each bubble is an $S^2$ rather than a circle, but the
``axis" connecting the two bubbles is still topologically $S^2$. The evolution
of this $S^2$ should be the same as before: it collapses down to a curvature
singularity in the future and past.

 The main difference is with the
behavior of the event horizon. This can be seen most easily by
considering the plane of symmetry
exactly halfway between the two bubbles. The induced metric is a four
dimensional spacetime, but there is one compact direction, so we can
think of this as a $2+1$ spacetime with $SO(2,1)$ symmetry. The metric
thus depends nontrivially on only one direction. The
singularity lies along a spacelike hyperboloid, and the event horizon is
the light cone of the origin. This is not a Killing horizon as we had
above. There is no killing field which is everywhere timelike outside the
horizon and becomes null on the horizon. The full event horizon is the
product of this light cone and a constant area two sphere (the analog of 
$\Sigma$). As a result, the area of the event
horizon grows with time and becomes arbitrarily large at infinity.
The topology of the horizon is $S^1 \times S^2$, so this is really a black 
string.
Since the singularity is along a spacelike hyperboloid in $2+1$ dimensions,
it intersects null infinity in a circle. This is consistent with the fact that
each bubble intersects null infinity in an $S^2$, and these two $S^2$'s 
intersect in a circle. The generalization to more than five dimensions
should be similar.

There are many open questions about the quantum properties of the black holes
that are produced by colliding bubbles. Presumably they emit Hawking radiation,
but how much energy can be emitted? The bubble spacetimes have zero
total energy initially, so in some sense the black hole has zero total
mass. It is known that there exists smooth initial data with the
topology of a bubble and arbitrarily negative energy \cite{Brill:qe}.
Could the black hole
radiate an arbitrarily large amount of energy? This may be difficult to address
in $3+1$ dimensions (where we have the exact classical solution),
since energy has unusual properties 
when there are only two noncompact spatial dimensions. We may 
have to wait for a better understanding of the higher dimensional
black holes formed from bubble collisions before one can answer this question.

\vskip 2in
\centerline{\bf Acknowledgements}
It is a pleasure to thank G. Gibbons, S. Giddings and E. Silverstein
for discussions. G.~H. wishes to thank the Isaac Newton Institute, Cambridge,
for hospitality while this work was being finished.
G.~H. is supported in part by NSF grant PHY-0070895. 
K.~M. is supported by a JSPS Postdoctoral Fellowship for 
Research Abroad.


\begin{thebibliography}{99}

\bibitem{Witten}
E. Witten, ``Instability Of The Kaluza-Klein Vacuum''  
Nucl. Phys. {\bf B195}, 481 (1982).

%\cite{Dowker:1995gb}
\bibitem{Dowker:1995gb}
F.~Dowker, J.~P.~Gauntlett, G.~W.~Gibbons and G.~T.~Horowitz,
``The Decay of magnetic fields in Kaluza-Klein theory,''
Phys.\ Rev.\ D {\bf 52}, 6929 (1995)
[arXiv:hep-th/9507143];
%%CITATION = HEP-TH 9507143;%%
``Nucleation of $P$-Branes and Fundamental Strings,''
Phys.\ Rev.\ D {\bf 53}, 7115 (1996)
[arXiv:hep-th/9512154].
%%CITATION = HEP-TH 9512154;%%

%\cite{Gutperle:2001mb}
\bibitem{Gutperle:2001mb}

J.~G.~Russo and A.~A.~Tseytlin,
``Magnetic flux tube models in superstring theory,''
Nucl.\ Phys.\ B {\bf 461}, 131 (1996)
[arXiv:hep-th/9508068];
%%CITATION = HEP-TH 9508068;%%
``Magnetic backgrounds and tachyonic instabilities in closed superstring  theory and M-theory,''
Nucl.\ Phys.\ B {\bf 611}, 93 (2001)
[arXiv:hep-th/0104238];
%%CITATION = HEP-TH 0104238;%%

M.~S.~Costa and M.~Gutperle,
``The Kaluza-Klein Melvin solution in M-theory,''
JHEP {\bf 0103}, 027 (2001)
[arXiv:hep-th/0012072];
%%CITATION = HEP-TH 0012072;%%

M.~Gutperle and A.~Strominger,
``Fluxbranes in string theory,''
JHEP {\bf 0106}, 035 (2001)
[arXiv:hep-th/0104136];
%%CITATION = HEP-TH 0104136;%%

S.~P.~De Alwis and A.~T.~Flournoy,
``Closed string tachyons and semi-classical instabilities,''
arXiv:hep-th/0201185.
%%CITATION = HEP-TH 0201185;%%

\bibitem{clean}
O. Aharony, M. Fabinger, G. Horowitz, E. Silverstein, 
``Clean Time-Dependent String Backgrounds from Bubble Baths'' 
arXiv:hep-th/0204158.  

%\cite{Birmingham:2002st}
\bibitem{Birmingham:2002st}
D.~Birmingham and M.~Rinaldi,
``Bubbles in anti-de Sitter space,''
arXiv:hep-th/0205246;
%%CITATION = HEP-TH 0205246;%%

V.~Balasubramanian and S.~F.~Ross,
``The dual of nothing,''
arXiv:hep-th/0205290;
%%CITATION = HEP-TH 0205290;%%

A.~M.~Ghezelbash and R.~B.~Mann,
``Nutty bubbles,''
arXiv:hep-th/0207123.
%%CITATION = HEP-TH 0207123;%%

%\cite{Fabinger:2000jd}
\bibitem{Fabinger:2000jd}
M.~Fabinger and P.~Horava,
``Casimir effect between world-branes in heterotic M-theory,''
Nucl.\ Phys.\ B {\bf 580}, 243 (2000)
[arXiv:hep-th/0002073];
%%CITATION = HEP-TH 0002073;%%

J.~Khoury, B.~A.~Ovrut, N.~Seiberg, P.~J.~Steinhardt and N.~Turok,
``From big crunch to big bang,''
Phys.\ Rev.\ D {\bf 65}, 086007 (2002)
[arXiv:hep-th/0108187].
%%CITATION = HEP-TH 0108187;%%

%\cite{Moss:iq}
\bibitem{Moss:iq}
I.~G.~Moss,
``Singularity Formation From Colliding Bubbles,''
Phys.\ Rev.\ D {\bf 50}, 676 (1994).
%%CITATION = PHRVA,D50,676;%%

%\cite{Emparan:2001wk}
\bibitem{Emparan:2001wk}
R.~Emparan and H.~S.~Reall,
``Generalized Weyl solutions,''
Phys.\ Rev.\ D {\bf 65}, 084025 (2002)
[arXiv:hep-th/0110258].
%%CITATION = HEP-TH 0110258;%%

\bibitem{Myers}
W. Israel and K. Khan, ``Collinear Particles and Bondi Dipoles in 
General Relativity", Nuovo Cimento, {\bf 33} 331 (1964);

R. Myers, ``Higher Dimensional Black Holes 
In Compactified Space-Times''  
Phys. Rev. {\bf D35}, 455 (1987).  

\bibitem{Einstein}
A. Einstein and N. Rosen, ``Two-body problem in general relativity
theory", Phys. Rev. {\bf 49}, 404 (1936).

\bibitem{kerr}
D. Kramer and G. Neugebauer, ``The superposition of two Kerr solutions",
Phys. Lett. {\bf 75A}, 259 (1980).

%\cite{Brill:qe}
\bibitem{Brill:qe}
D.~Brill and H.~Pfister,
``States Of Negative Total Energy In Kaluza-Klein Theory,''
Phys.\ Lett.\ B {\bf 228}, 359 (1989).
%%CITATION = PHLTA,B228,359;%%

D.~Brill and G.~T.~Horowitz,
``Negative Energy In String Theory,''
Phys.\ Lett.\ B {\bf 262}, 437 (1991).
%%CITATION = PHLTA,B262,437;%%

\end{thebibliography}
\end{document}